\documentstyle[emulateapj,psfig]{article}
\begin{document}    
\newcommand{\be}{\begin{equation}}
\newcommand{\ba}{\begin{eqnarray}}
\newcommand{\ee}{\end{equation}}
\newcommand{\ea}{\end{eqnarray}}  
       
\title{The Influence of Galactic Outflows on
the Formation of Nearby Dwarf Galaxies}

\author{Evan Scannapieco\altaffilmark{1}, Andrea Ferrara\altaffilmark{2}, 
\& Tom Broadhurst\altaffilmark{3}}
      
\altaffiltext{1}{
Department of Astronomy,
University of California, Berkeley, CA 94720}
\altaffiltext{2}{Osservatorio Astrofisico di Arcetri
        50125 Firenze, Italy}
\altaffiltext{3}
{European Southern Observatory, Karl-Schwarzschild-Strasse 2,
	Garching bei M\"{u}nchen, Germany}

\begin{abstract}

We show that the gas in growing density perturbations is vulnerable to
the influence of winds outflowing from nearby collapsed galaxies
that  have already formed stars.  This suggests that 
the formation of nearby galaxies with masses $\lesssim 10^9 M_\odot$
is likely to be suppressed, irrespective of the details of galaxy formation.  
An impinging wind may shock heat
the gas of a nearby perturbation to above the virial temperature,
thereby mechanically evaporating the gas, or the baryons may be stripped
from the perturbation
entirely if they are accelerated to above the escape velocity.  
We show that baryonic stripping is the most effective of these two processes,
because shock-heated clouds that are too large to be stripped 
are able to radiatively cool within a sound-crossing time, limiting 
evaporation.  The intergalactic medium
 temperatures and star-formation rates required 
for outflows to have a significant influence on the formation of
low-mass galaxies are consistent with current observations, but may soon be
examined directly via associated distortions in the cosmic
microwave background, and with near-infrared observations from the Next
Generation Space Telescope, which may detect the supernovae 
from early-forming stars.

\end{abstract}

\section{Introduction}

Outflows from dwarf starbusting galaxies have long been the subject of
theoretical and observational investigation.  Since the 1970s,
theoretical work has shown that supernovae (SNe) and OB winds in these
low-mass objects produce energetic outflows that
shock and enrich the intergalactic medium (IGM) in which
they are embedded (Larson 1974; Dekel \& Silk 1986; Vader 1986).  This
behavior has been clearly identified in studies of both local
starbursting galaxies (Axon \& Taylor 1978; 
Heckman 1997; Martin 1998)
and their environments (della Ceca et al.\ 1996; Bomans, Chu, \& Hopp
1997) as well as with spectroscopy of high-$z$ galaxies (Franx et al.\
1997; Pettini et al.\ 1998; 
Frye, Broadhurst \& Sprinrad 1999).

Whether these outflows lead to a catastrophic loss of the interstellar
gas in these objects is much more uncertain, with recent
simulations and observations suggesting that dwarfs are very
inefficient in removing gas from their cores (Mac Low \& Ferrara 1999;
Murakami \& Babul 1999).  However, the issue of whether 
dwarfs retain a sizeable fraction of
their gas is to some degree decoupled from the formation of outflows.
In galaxies in the mass range $10^7 M_\odot \lesssim M \lesssim 10^9
M_\odot$, Mac Low \& Ferrara (1999) have shown that a ``blowout''
occurs in which the superbubbles produced by multiple SNII explosions
punch out of the galaxy, shocking the surrounding IGM while failing to
excavate the interstellar medium of the galaxy as a whole.

In the generally investigated hierarchical models of structure
formation, low-mass galaxies form in large numbers at
early times (e.g. White \& Frenk 1991) are highly clustered (Kaiser
1984).  The existence of a large number of small and clustered
galaxies at early times is also favored observationally by the steep
number counts and low luminosities of faint galaxies (Broadhurst,
Ellis \& Glazebrook 1992), the sizes of faint galaxies in the Hubble
Deep Field (HDF) (Bouwens, Broadhurst, \& Silk 1999a,b), and the
clustering properties of Lyman-break galaxies (Adelberger et al.\ 1998).
Locally however, the number density of low-mass objects is far less than
predicted theoretically (Ferguson \& Binggeli 1994; Klypin et al.\
1999; Moore et al.\ 1999), leading to theoretical studies
into the disruption of dwarves by tidal forces from neighboring
objects (Moore, Lake, \& Katz 1998) and external UV radiation (
Norman \& Spaans 1997; Corbelli, Galli, \& Palla 1997; 
Ferrara \& Tolstoy 2000).

Little attention has been directed however, toward the
influence of outflows on the formation of neighboring galaxies, 
despite the fact that
typical outflow velocities exceed the virial
velocities of objects in the mass range in which the
overabundance is most severe.
In a companion article (Scannapieco \& Broadhurst 2000) we describe
the effects of heating and enrichment by outflows on galaxy
formation using a Monte Carlo treatment of hierarchical structure formation.
Here we show using simple scaling arguments that suppression of
galaxy formation by outflows from dwarf galaxies is important over a
large range of halo masses, irrespective of the details of
cosmological simulations. 

The structure of this work is as follows.  In \S2 we consider two
scenarios for the suppression of dwarf galaxy formation by outflows.
In \S3 we show that the outflow models considered fall well within the 
bounds of current observations.  Conclusions are given in \S4.

\section{Suppression of Galaxy Formation}

We consider two processes by which outflows from neighboring dwarves
inhibit the formation of a galaxy. In the 
``mechanical evaporation'' scenario, the gas associated
with an overdense region is heated by a shock to above its 
virial temperature. The thermal pressure of the gas then
overcomes the dark matter potential and the gas expands out of the
halo, preventing galaxy formation.  In this case, the cooling
time of the collapsing cloud must be shorter than its sound crossing
time, otherwise the gas will cool before it
expands out of the gravitational well and
will continue to collapse.

Alternatively, the gas may be stripped from a perturbation by a shock
from a nearby source.  In this case, the momentum of the 
shock is sufficient to carry with it the gas associated with the
neighbor, thus emptying the halo of its baryons and preventing a galaxy from 
forming. In the following we evaluate the importance of these two effects
in turn.

\subsection{Mechanical Evaporation}

The first mechanism we consider is the shock heating of the halo
gas to a temperature sufficient to cause the gas to evaporate into
the IGM.  This will happen for cases in which the virial temperature 
of the object $T_v$ is less than the postshock temperature $T_s$,
where $T_s$ can be written for an adiabatic shock as 
\be
T_s = {3 \mu m_p\over 16 k} v_s^2 = 14 \, v_{s,km/s}^2 \, {\rm K},
\label{eq:temp}
\ee
where $\mu$ is the molecular weight of the gas, which we take to be 0.6, 
 $m_p$ is the proton mass, $k$ is Boltzmann's constant, and
$v_s$ is the velocity of the expanding shock.

For the typical outflows we will be considering, the blast wave
reaches the neighboring perturbation of interest within one
cooling time $t_c \approx \frac{k T_s}{\bar{n}_e  \Lambda}$, where 
$\bar{n}_e = 4.5 \times 10^{-7} (1+z)^3 h^2 \, {\rm cm}^{-3}$
is the average number density of electrons, 
$\Lambda$ is the cooling function
which we normalize to $\Lambda_{-22} \equiv \Lambda/10^{-22}$ 
erg~cm$^{3}$~s$^{-1}$, and $h$ is the Hubble constant in units
of 100 km/s/Mpc.   Here and below we take the baryonic density 
in units of the critical density to be $\Omega_b = 0.05$. Thus the 
shock velocity can be approximated by a Sedov-Taylor blast wave solution,
\be
v_{s} = 1600 \, (\epsilon{\cal N} h)^{1/2} \, d_{c, {\rm kpc}}^{-3/2} \, 
{\rm km/s},
\label{eq:vs}
\ee
where ${\cal N}$ is the number of supernovae driving the bubble (each 
with a total energy $2\times 10^{51}$ erg to take into account the contribution
from stellar winds), $\epsilon$ is the total-to-kinetic energy conversion 
efficiency, $d_{c,{\rm kpc}}$ is the comoving distance from the explosion site
in kpc $\times h$. The Compton drag on the expanding shock
as well as the Hubble expansion are easily shown to be negligible
for our purposes in this paper (Ferrara 1998; Scannapieco \& Broadhurst 2000).
From the relation $T_v = 70 M_{6}^{2/3}(1+z) \Omega_0^{1/3}$~K,
where $M_{6}\equiv M/(10^6 M_\odot/ h)$
and $\Omega_0$ is the matter density in units of the
critical density,
it follows that 
collapsed objects with total masses lower than
$M_6 \leq 3.5 \times 10^{8} (1+z)^{-3/2} (\epsilon {\cal N} h) ^{3/2} 
 d_{c,{\rm kpc}}^{-9/2} \Omega_0^{-1/2} 
$
will be affected by shock heating and evaporation.
We estimate $d_{c,{\rm kpc}}$ as the mean spacing between objects of 
a mass scale $M$, which is appropriate as the outflowing objects and the
forming density perturbations of interest are of roughly the
same mass scales:
\be
d_{c,{\rm kpc}} \approx \left({3 M\over {4\pi} \rho_c \Omega_0}\right)^{1/3} 
\simeq  10 M_6^{1/3} \Omega_0^{-1/3}
, \label{eq:meansep}
\ee
which gives
\be
M_6 \lesssim 41 (1+z)^{-3/5} (\epsilon {\cal N} h)^{3/5} 
\Omega_0^{2/5}.
\label{eq:evap}
\ee 

Assuming that one SN occurs for every 100 $M_\odot$ of baryons which
form stars (see eg.\ Gibson 1997) we can relate the number of
SNe and the mass of the halo as
${\cal N} h   = 500 M_6  \epsilon_{sf} \, \Omega_0^{-1},$
where $\epsilon_{sf}$ is the initial star formation efficiency.
Assuming an initial star formation efficiency of $\epsilon_{\rm sf} = 0.1$
and a kinetic energy conversion efficiency of $\epsilon = 0.2$ we can
estimate $\epsilon {\cal N} h$ as approximately $10 M_6 \Omega_0^{-1}$, 
and thus we  expect $\epsilon {\cal N} h \approx 5000 \Omega_0^{-1}$ for 
typical dwarf-galaxy sized objects of $ \sim 5 \times 10^8 M_\odot.$

The relation given by Eq. (\ref{eq:evap}) is shown by the upper
lines in Fig.\ 1
for $\epsilon {\cal N} h = 10^4$.  Also
shown for reference on this plot is the mass below which
effects due to photoevaporation are important ($M_{6} \lesssim 4.4\times 10^3
(1+z)^{-3/2} \Omega_0^{-1/2}$). Galaxies with masses below this limit are
readily evaporated by the UV background after the reionization of the
universe (Barkana \& Loeb 1999;  Ferrara \& Tolstoy 2000)

\subsubsection{Cooling}

To suppress galaxy formation, however, it is not enough only to
heat the halo gas,  because the gas could cool rapidly and fall back
towards the center of the object.
Therefore Eq.\ (\ref{eq:evap}) 
is a necessary but not sufficient condition for mechanical evaporation.
The second condition requires that the gas remains hot for the time
required to escape the system, or that the cooling time be longer than the
the sound crossing time, ${kT_s\over n_e \Lambda} \ge 
{\ell / c_s}$, where $\ell$ is size of the halo,
 $c_s=(kT_s/\mu m_h)^{1/2}$ is the sound speed, and
$n_e = \bar{n}_e \delta $  where 
$\delta \equiv \rho/\rho_0$ is the overdensity of the 
region, $\rho_0$ is the mean density of the universe.
This gives
\begin{eqnarray}
M_6 \lesssim 
2.3 (\epsilon N h)^{9/11} (1+z)^{-12/11} \delta^{-4/11} h^{-6/11}\nonumber\\ 
\times \Omega_0
\Lambda^{-6/11}_{-22},
\end{eqnarray}
which forms the upper bound of the cross-hatched region in Figure 1.
Here we see that for both cosmological models, the cooling vastly
undercuts the minimum mass necessary for disruption by heating,
dropping below the photoevaporation limit in the $\Lambda$CDM model.
While it is relatively easy for galactic outflows to shock low-density 
gas to above the virial temperatures of the potential, this excess 
energy is efficiently radiated away for most large clouds.  

\subsection{Baryonic Stripping}

The second possibility is that all the gas is striped by the impinging
shock.  We may estimate that 
this will occur when the shock moves through the center of the
halo with sufficient momentum to accelerate the gas to the escape velocity
of the halo: $f M_s v_s > M_c v_e$
where $f=\ell^2/4 d^2$ is the solid angle of the shell impinging on
the lump,  $M_s$ is the mass of material swept up by the expanding shock, 
$M_c=(4\pi/3)\rho_c \Omega_b \delta \ell^3$ is the baryonic mass of the 
cloud whose radius is $\ell$, and $v_e=(G M/\ell)^{1/2}$
is the escape speed from the cloud.  Solving for the mass that
can be swept up as function of redshift 
and replacing $d_{c,{\rm kpc}}$ by the mean separation as estimated from
Eq.\ (\ref{eq:meansep}) gives
\be
M_6 \lesssim 53  \delta^{-1} (1+z)^{-3/5} 
      (\epsilon {\cal N} h)^{3/5} \Omega_0^{2/5}.
\ee
which is also plotted in Figure 1.  Here we see that while it is
significantly easier to heat a cloud than to sweep away its
baryons, the short cooling times of most halos cause baryonic
stripping to have the greatest impact on the formation of nearby
galaxies.  Note that in this plot the density is taken to be only
twice that of the mean density of the universe.  
Thus it is important that the perturbation be in the 
linear or weakly nonlinear regime for this mechanism
to be effective.  As the time between turn-around and 
virialization is relatively short, however, most protogalaxies are likely
to be at these low density contrasts when shocked by neighboring objects.

\section{Observational Constraints}

\subsection{Compton y-parameter}

While widespread baryonic stripping of dwarf galaxies is a
generic consequence of associating outflows typical of local dwarfs
with primordial dwarf galaxies, it is important to check that this
extrapolation to higher redshift is consistent with observational constraints.
An unavoidable result of widespread IGM heating is the 
presence of spectral distortions in the cosmic microwave background
(CMB).  The degree of these distortions is given by the
Compton $y$ parameter which is simply the convolution of the optical depth
with the electron temperature along the line of sight
(Zel'dovich \& Sunyaev 1969; Sunyaev \& Zel'dovich 1972).
To calculate the mean $y$ we make use of the fact that the
total optical depth within ionized regions must be within the
observational limit of $\tau \lesssim 0.5$ (Griffiths, Barbosa, \&
Liddle 1999).
As the total optical depth within outflows must be less than the total
optical depth overall we can estimate this convolution as
\be
y \lesssim {\tau\over V_4} 
\left(\int_{r_{\rm i}}^{r_{10^4}} \frac{kT_s(r)}{m_e c^2} 4\pi r^2 dr \right),
\ee
where $r_i$ is the initial size of the blast, which we take
to be $1 h^{-1}$ comoving kpc, and $V_4$ is the volume defined by the radius 
outside which  the temperature of the blast drops below the ionization temperature of
hydrogen $\approx 10^4$, which is readily estimated from Eq.\ (\ref{eq:temp})
and (\ref{eq:vs}) to be equal to  $15 \, (\epsilon {\cal N} h)^{1/3} \,h^{-1}
\, (1+z)^{-1}$~kpc.

This gives
\be
y \lesssim \tau \frac{k 10^4 K}{m_e c^2} \ln(15 (\epsilon {\cal N} h)^{1/3})
    \lesssim 1.4 \times 10^{-5},
\ee
where we take $(\epsilon {\cal N} h) = 10^4$.  This is within the
bounds set by the COBE data of $y \leq 1.5 \times 10^{-5}$ (Fixsen et
al.\ 1996), but only marginally.  This suggests that future CMB experiments
may help to constrain the degree of IGM heating by outflows.

\subsection{Point Source Luminosities}

A second consequence of widespread dwarf outflows
is the presence of objects seen as point sources in deep surveys.
A SN number of ${\cal N} \approx 5 \times 10^4$ implies that the local star formation
rate is roughly $0.5
M_\odot$/yr.  The question arises if objects with these luminosities 
could have escaped detection as point sources in the 
HDF. 
Assuming a Salpeter initial mass function, Ciardi et al.\ (2000) give the 
luminosity per unit frequency at the Lyman limit obtained
from the adopted spectrum of a primordial galaxy at early evolutionary times
as  $j_0  = 4\times 10^{20} M_{\star} \; {\rm erg} \, {\rm s}^{-1} {\rm
Hz}^{-1}, $ where $M_\star$ is the stellar mass of the object.
Scaling to the stellar mass value implied by the assumed efficiency of massive
star formation, $M_{\star}=5\times 10^6 M_\odot$, we find for the luminosity of the SN parent cluster of stars
$L_\nu \simeq 6\times 10^{42}$~erg/s at the Lyman limit. 
The observed flux is then
${\cal F}(\nu_0) = L_\nu (1+z)/ 4\pi d_L^2,$
where $\nu_0=\nu/(1+z)$ is the observed frequency.
This  gives approximately ${\cal
F}=12 (1+z)^{-1} h^2$~nJy. In order to be observed in the HDF 
optical filter centered at 6060\AA\ (V$_{606}$), the object should be 
located at redshift $1+z=6.6$, thus with a flux equal to $\approx 
2 h^2$~nJy. This value corresponds to  
a AB magnitude $V_{AB}=30.6$ which 
is well below the limiting magnitude of the HDF for point sources
(typically 28 mag, see Mannucci \& Ferrara 1999). 
However, these objects may 
be detectable by the Next Generation Space Telescope, 
which should reach AB magnitudes of order 32 in the near infrared,
although dust extinction may complicate this analysis.

\section{Conclusions}

Using simple scaling arguments, we have shown that outflows from 
low-mass galaxies play an important role in the suppression of the
formation of nearby dwarf galaxies.  While the details
of this process depend on the spatial distribution of forming galaxies,
and can be only be studied in the context of cosmological simulations
(Scannapieco \& Broadhurst 2000), the overall implications of this mechanism
can be understood from the general perspective of hierarchical
structure formation.  

In principle, outflows can suppress the formation of nearby galaxies both by
shock heating and evaporation and by stripping of the baryonic matter
from collapsing dark mater halos; in practice, the short cooling times for most
dwarf-scale collapsing objects suggest that the baryonic stripping scenario is
almost always dominant.  This mechanism has the largest impact in
forming dwarves in the $\lesssim 10^9 M_\odot$ range which is sufficiently
large to resist photoevaporation by UV radiation, but too small to
avoid being swept up by nearby dwarf outflows. 

It is interesting to note that numerical studies working from a
different perspective have similarly shown that 
momentum transfer is the most important feedback mechanism
in determining  the internal structure of larger galaxies.
While SN heating is relatively ineffective at regulating star formation
in numerical models, accounting for the kinetic component
of SN feedback has resulted in quasi-stationary models of disc galaxies
with reasonable star formation rates (see eg.\ Katz 1992;
Navarro \& White 1993; Mihos \& Hernquist 1994; Springel 2000).

Various analytical and N-body studies (Kauffmann, White, \& Guiderdoni
1993; Klypin et al.\ 1999; Moore et al.\ 1999) have shown that $\sim
50$ satellites with circular velocities $\sim 20$ km/s and masses
$\sim 10^9 M_\odot$ should be found within 600 kpc of the Galaxy,
while only 11 are observed.  While the fraction of objects suppressed
is dependent on the assumed cosmology and spatial distribution of galaxies,
the partial suppression of the formation of objects in this mass range
is a general consequence of baryonic stripping by outflows.
Note that dark matter halos which are subject to sweeping are
likely to accrete some gas at later times and thus
this scenario provides a natural mechanism for the formation 
of ``dark'' Milky Way satellites, which may be
associated the abundant High-Velocity Clouds as discussed by Blitz
et al.\ (1999).  Also, this population of halos
could be identified with the low mass tail distribution of the dark 
galaxies that reveal their presence through gravitational lensing of 
quasars (Hawkins 1997).

The existence of an era of widespread IGM heating through outflows
is consistent with current CMB constraints, but causes a mean
spectral distortion in the range that will be probed by the
next generation of experiments.  Similarly, the luminosities of the
majority of the starbursting dwarfs that contribute to this process
are beyond the limiting magnitudes of the Hubble Deep Field, but may be 
observable with of the Next Generation Space
Telescope.  Thus it may be in the near future that we can directly
observe the era in which dwarf galaxies were first formed and the 
impact of galactic outflows on this process.

\acknowledgments

We would like to thank Marc Davis, Joseph Silk, and an anonymous
referee  for helpful comments and discussions.  This research was 
supported in part by the National Science Foundation under Grant 
PHY94-07194.
 
\fontsize{9}{11pt}\selectfont

\begin{figure*}
\centerline{\psfig{file=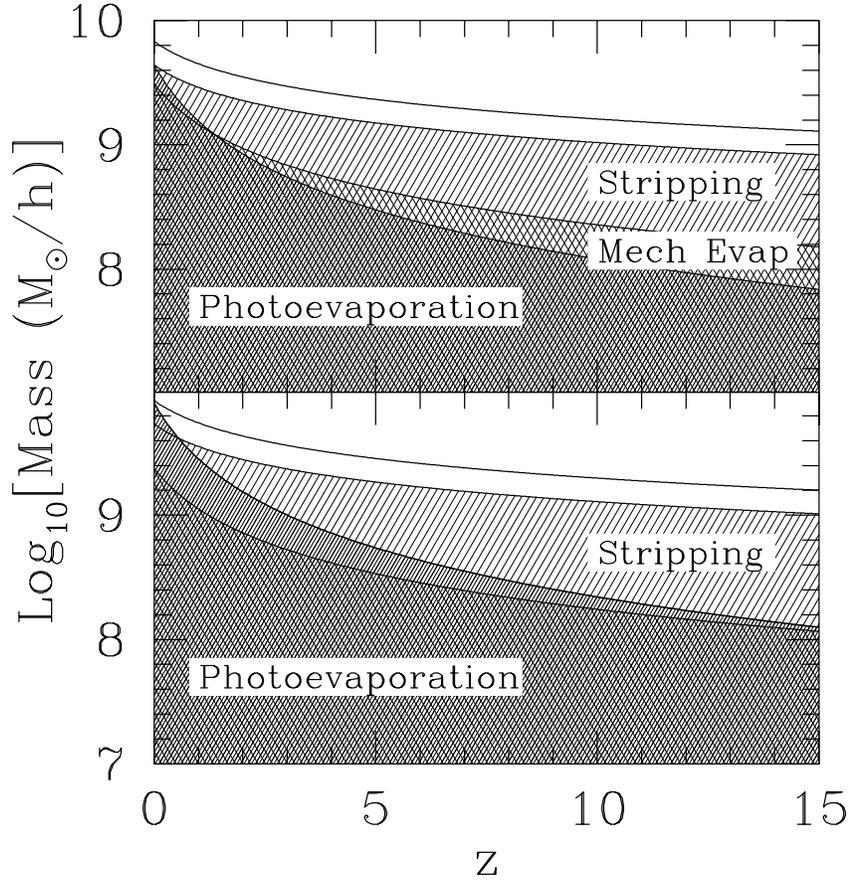,width=14 cm} }
\caption{\footnotesize{
Relevant mass scales for suppression of dwarf galaxy formation.
The upper lines are the masses
below which halos will be heated beyond their virial temperatures,
although cooling prevents mechanical evaporation from occurring for halos
with masses above the cross hatched regions.  The second highest set of
lines, bounding the lightly shaded regions, show the masses below which
baryonic stripping is effective.   Finally the 
heavily shaded regions show objects that are susceptible to photoevaporation.
The upper panel 
is a flat CDM model and the lower panel is a flat $\Lambda$CDM model
with $\Omega_0 = 0.3$.  In all cases $(\epsilon N h) = 5000 \Omega_0^{-1}$,
the overdensity $\delta = \rho/\rho_0 = 2.0$, $\Omega_b = 0.05$, $h = .65$,
and the cooling function $\Lambda_{-22} = 1$ as described in \S 2.1.1.
Note that photoevaporation affects a larger mass range than mechanical
evaporation in the $\Lambda$CDM cosmology.}}
\label{fig:etacompare}
\end{figure*}


\begin{references}
\reference{} Adelberger, K. L., Steidel, C. C., Giavalisco, M. 
	     Dickinson, M., Pettini, M., \& Kellogg, M. 1998,
	      ApJ, 505, 18
\reference{} Axon, D. J., \& Taylor, K. 1978, Nature, 274, 37
\reference{} Barkana, R. \& Loeb, A. 1999, ApJ, 523, 54
\reference{} Blitz, L., Spergel, D. N., Teuben, P. J., Hartmann, D., Burton, 
	     W. B. 1999, ApJ, 514, 818
\reference{} Bomans, D. J., Chu, Y.-H., \& Hopp, U. 1997, AJ, 113, 167
\reference{} Broadhurst, T., Ellis, R., \& Glazebrook, K. 1992, 
         Nature, 355, 55
\reference{} Bouwens, R., Broadhust, T., \& Silk, J. 1998a, ApJ, 506, 557
\reference{} Bouwens, R., Broadhust, T., \& Silk, J. 1998b, ApJ, 506, 579
\reference{} Ciardi, B., Ferrara, A., Governato, F., \& Jenkins, A. 2000,
        MNRAS, in press, (astro-ph/9907189)
\reference{} Corbelli, E., Galli, D., \& Palla, F. 1997, 487, L53
\reference{} Dekel, A., \& Silk, J. 1986, ApJ, 303, 39
\reference{} della Ceca, R., Griffiths, R. E., Heckman, T. M., \& MacKenty,
        J. W. 1996, ApJ, 469, 662
\reference{} Ferguson, H. C., \& Binggeli, B. 1994, A\&AR, 6, 67
\reference{} Ferrara, A. 1998, ApJ, 499, L17
\reference{} Ferrara, A., \& Tolstoy, E. 2000, MNRAS, 313, 291 
\reference{} Fixsen, D. J., Cheng, E. S., Gales, J. M., Mather, J. C., 
        Shaver R. A., \& Wright, E. L. 1996, ApJ, 473, 576
\reference{} Franx, M., Illingworth, G. D., Kelson, D. D., van Dokkum, P. G.,
             \& Tran, K.-V. 1997, ApJ, 486, L75
\reference{} Frye, B., Broadhurst, T., \& Spinrad, H. 1999, 
        in Gravitational Lensing: Recent Progress and Future Goals, 
        Boston University, 25-30 July 1999, to be published.
\reference{} Gibson, B. K. 1997, MNRAS, 290, 471
\reference{} Griffiths, L. M., Barbosa, D., \& Liddle, A. R. 
        1999, MNRAS, 308, 854
\reference{} Hawkins, M. R. S. 1997, A\&A, 328, L25
\reference{} Heckman, T. 1997, RevMexAA, 6, 156
\reference{} Kaiser, N. 1984, ApJ, 284, L9
\reference{} Katz, N. 1992, ApJ, 391, 502
\reference{} Kauffmann, G., White, S. D. M., \& Guiderdoni, B. 1993, 
        MNRAS 264, 201
\reference{} Klypin, A., Kravtsov, A. V., Valenzuela, O., \& Prada,
        F. 1999, ApJ, 522, 82
\reference{} Larson, R. B., 1974, MNRAS, 166, 585
\reference{} Mac Low, M. M., \& Ferrara, A. 1999, ApJ, 513, 142
\reference{} Mannucci, F., \& Ferrara, A. 1999, MNRAS, 305, 55
\reference{} Martin, C. L. 1998, ApJ, 506, 222
\reference{}Mihos, J. C., \& Hernquist, L. 1994, ApJ, 427, 611
\reference{} Moore, B., Lake, G., \& Katz, N. 1998, ApJ, 495, 139
\reference{} Moore, B., Ghinga, F., Governato, G., Lake, T., Quinn,
        T., Stadel, J., \& Tozzi, P. 1999, ApJ, L19
\reference{} Murakami, I., \& Babul, A. 1999, MNRAS, 309, 16
\reference{}Navarro, J. F., \& White, S. D. M. 1993, MNRAS, 265, 271
\reference{} Norman, C. A., \& Spaans, M. 1997, ApJ, 480, 145
\reference{} Pettini,  M., Kellogg, M., Steidel, C. C., Dickinson, M. E.,         Adelberger, K. L., \& Giavalisco, M. 1998, ApJ, 508, 539 
\reference{} Scannapieco, E., \& Broadhurst, T. 2000, ApJ submitted,
        (astro-ph/0003104)
\reference{}Springel, V. 2000, MRNAS, 312, 859
\reference{} Sunyaev, R. A., \& Zel'dovich, Ya. B. 1972, Comm. 
         Astrophys. Space Phys., 4, 173 
\reference{} Vader, J. P. 1986, ApJ, 305, 669
\reference{} White, S. D. M., \& Frenk, C. S. 1991, ApJ, 379, 52
\reference{} Zel'dovich Ya. B., \& Sunyaev, R. A. 1969, Ap. Space Sci., 
        4, 301
\end{references}
\end{document}